\documentclass{article}
\usepackage{spconf,amsmath,graphicx}
\usepackage{hyperref}
\usepackage{caption}
\usepackage{algorithm,algorithmicx,algpseudocode}

\usepackage{amsmath,amsfonts,amssymb,pifont}
\usepackage{adjustbox}
\usepackage{comment}

\title{High-Fidelity Speech Synthesis with Minimal Supervision: All Using Diffusion Models}

\name{Chunyu Qiang$^{1,2,3}$, Hao Li$^{2}$, Yixin Tian$^{2}$, Yi Zhao$^{2}$, Ying Zhang$^{2}$, Longbiao Wang$^{1,3,*}$, Jianwu Dang$^{3}$}

\address{$^1$School of New Media and Communication, Tianjin University, Tianjin, China \\
$^2$Kuaishou Technology Co., Ltd, Beijing, China \\
$^3$Tianjin Key Laboratory of Cognitive Computing and Application, \\
College of Intelligence and Computing, Tianjin University, Tianjin, China }

\begin{document}
\ninept
\maketitle

\begin{abstract}
Text-to-speech (TTS) methods have shown promising results in voice cloning, but they require a large number of labeled text-speech pairs. Minimally-supervised speech synthesis decouples TTS by combining two types of discrete speech representations(semantic \& acoustic) and using two sequence-to-sequence tasks to enable training with minimal supervision. However, existing methods suffer from information redundancy and dimension explosion in semantic representation, and high-frequency waveform distortion in discrete acoustic representation. Autoregressive frameworks exhibit typical instability and uncontrollability issues. And non-autoregressive frameworks suffer from prosodic averaging caused by duration prediction models. To address these issues, we propose a minimally-supervised high-fidelity speech synthesis method, where all modules are constructed based on the diffusion models. The non-autoregressive framework enhances controllability, and the duration diffusion model enables diversified prosodic expression. Contrastive Token-Acoustic Pretraining (CTAP) is used as an intermediate semantic representation to solve the problems of information redundancy and dimension explosion in existing semantic coding methods. Mel-spectrogram is used as the acoustic representation. Both semantic and acoustic representations are predicted by continuous variable regression tasks to solve the problem of high-frequency fine-grained waveform distortion. Experimental results show that our proposed method outperforms the baseline method. We provide audio samples on our website. \href{https://qiangchunyu.github.io/Diff-Speech/}{$^1$}
\end{abstract}

\renewcommand{\thefootnote}{\fnsymbol{footnote}} 
\footnotetext{$*$ Corresponding author.} 
\footnotetext{Audio samples: https://qiangchunyu.github.io/Diff-Speech/}

\begin{keywords}
minimal supervision, speech synthesis, semantic coding, diffusion model, CTAP
\end{keywords}
\section{Introduction}
\label{sec:intro}
With the development of deep learning, existing speech synthesis methods have achieved satisfactory results \cite{wang2017tacotron,arik2017deep, li2019neural,ren2019fastspeech, kim2020glow, elias2021parallel}. Traditional speech synthesis methods typically use mel-spectrograms as an intermediate representation. However, recent advances in neural vocoders have led TTS methods to convert audio waveforms into discrete codes as an intermediate representation \cite{baevski2020wav2vec, Hsu2021HuBERTSS, Defossez2022HighFN, Zeghidour2022SoundStreamAE}. These TTS systems can be roughly divided into two categories: 1) autoregressive frameworks \cite{borsos2022audiolm, wang2023neural,zhang2023speak,kharitonov2023speak} and 2) non-autoregressive frameworks \cite{levkovitch2022zero, shen2023naturalspeech, le2023voicebox}, both of which aim to voice cloning. For instance, VALL-E\cite{wang2023neural} is the first large-scale TTS framework based on a language model with in-context learning capabilities for zero-shot speech synthesis. Naturalspeech2 \cite{shen2023naturalspeech} is a non-autoregressive TTS framework based on a latent diffusion model \cite{ho2020denoising}.  Although these methods can achieve good voice cloning effects, they rely on a large number of labeled text-speech pairs. 

SPEAR-TTS\cite{kharitonov2023speak} is another example, which divides the TTS task into two tasks: text-to-semantic (supervised learning) and semantic-to-speech (unsupervised learning), using discrete semantic coding (Wav2Vec-BERT\cite{chung2021w2v}) and acoustic coding (SoundStream\cite{Zeghidour2022SoundStreamAE}) as intermediate representations to achieve minimally-supervised training. However, discrete acoustic coding relies on neural vocoders for speech waveform reconstruction, and there is information loss in high-frequency fine-grained acoustic details compared to traditional audio features. The information content of the semantic coding is expected to be a "bridge" between text and acoustic information. It should emphasize linguistic content while de-emphasizing paralinguistic information such as speaker identity and acoustic details. However, the semantic coding extracted by existing models suffers from excessive redundancy and dimension explosion, leading to difficulties in prediction from text and cumulative errors. Additionally, the autoregressive framework suffers from the typical problems of instability and uncontrollability. These issues have been verified in our previous work\cite{qiang2023minimally}. To address these problems, we propose a minimally-supervised high-fidelity speech synthesis method based entirely on diffusion models. The main contributions include: 

\begin{figure*}[t]
 \centering
 \includegraphics[width=\linewidth]{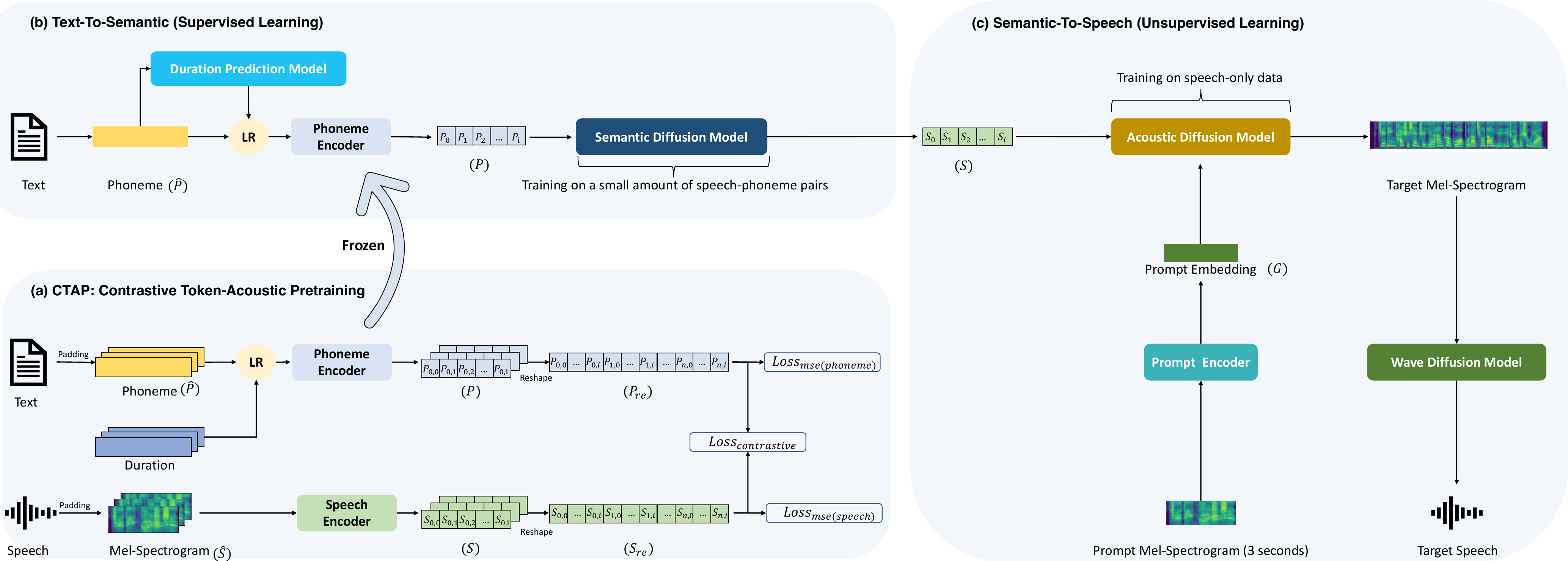}
 \vspace{20pt}
 \captionsetup{belowskip=20pt}
 \caption{In the Text-To-Semantic, supervised training is performed using speech-phoneme pairs. In the Semantic-To-Speech, unsupervised training is performed using speech-only data. The CTAP\cite{qiang2023cpsp} jointly trains a speech encoder and a phoneme encoder using contrastive learning to learn frame-level (dis)similarity between a batch of speech-phoneme pairs. The pre-trained phoneme encoder and speech encoder are frozen to extract phoneme representation $P$ and speech representation $S$.}
 \label{fig:proposed_model}
\end{figure*}

1) We propose a minimally-supervised speech synthesis method, where all modules are constructed based on diffusion models. The non-autoregressive framework achieves better controllability, and the duration diffusion model achieves diversified prosodic expression, solving the problem of expression averaging caused by the duration prediction model. 

2) We introduce our previously proposed speech coding method CTAP\cite{qiang2023cpsp}\cite{qiang2023ctap} as an intermediate semantic representation to solve the problem of information redundancy and dimension explosion in existing semantic coding methods. We also introduce a prompt encoder structure based on a variational autoencoder and a prosody bottleneck to improve prompt representation ability. 

3) The mel-spectrograms are used as acoustic representations. Both semantic representation and acoustic representation prediction are continuous variable regression tasks, solving the problem of high-frequency fine-grained waveform distortion.

\section{Method}
\subsection{Overview}

Minimally-supervised TTS involves splitting the TTS into two tasks (text-to-semantic and semantic-to-speech) by combining speech intermediate representations. As shown in Fig. \ref{fig:proposed_model}, extends SPEAR-TTS\cite{zhang2023speak} by enabling the diffusion model for non-autoregressive continuous variable regression tasks.

For the text-to-semantic task, as shown in Fig. \ref{fig:proposed_model}(b), supervised training is performed using labeled speech-phoneme pairs. Phonemes are upsampled according to duration via the length regulator. During training, ground truth duration is used to extend the phoneme sequence, while the corresponding predicted duration is used during inference. Then, the frozen phoneme encoder is input to obtain phoneme representation $P$, which is then input to the semantic diffusion model to predict speech representation $S$. The phoneme and speech representations extracted by the CTAP model based on contrastive learning are very similar, making the prediction task of the semantic diffusion model simple. 

For the semantic-to-speech task, as shown in Fig. \ref{fig:proposed_model}(c), unsupervised training is performed using unlabeled speech-only data. The acoustic diffusion model maps semantic representation $S$ to mel-spectrograms, while prompt embeddings provide paralinguistic information from the prompt speech, such as timbre, style, and prosody. The wave diffusion model is then used to convert mel spectrograms into speech.

\subsection{Prompt Encoder}

The prompt encoder is a VAE-based model \cite{qiang2022style} that extracts paralinguistic information from the prompt speech, such as timbre, style, and prosody. It comprises a 6-layer 2D convolutional network and a SE-ResNet block \cite{hu2018squeeze}, which recalibrates channel-wise feature responses by modeling interdependencies among channels, resulting in significant performance improvements. The VAE structure enables the model to obtain a continuous and complete latent space distribution of styles, improving the ability to extract paralinguistic information. A 64-dimensional vector is sampled from the Gaussian distribution as the prompt embedding. At each step, we randomly clip the prompt speech with a window of length 3 seconds to use as input for the prompt encoder. The prompt embedding is denoted by $G$. To address the KL collapse problem, three tricks are used: 1) introducing KL annealing, 2) adopting a staged optimization method to optimize the reconstruction loss first and then the KL loss, and 3) A margin $\Delta$ is introduced to limit the minimum value of the KL loss as shown:
$\mathcal{L}_{kl} = max(0, D_{KL}[\mathcal{N}({\hat{\mu}},{\hat{\sigma}}^2)||\mathcal{N}(0, I)]-\Delta)$.

\subsection{CTAP: Contrastive Token-Acoustic Pretraining}

CTAP is illustrated in Fig. \ref{fig:proposed_model}(a). CTAP mainly consists of four components, including a speech encoder, a phoneme encoder, a prompt encoder, and a decoder. The length regulator is used to solve the problem of length mismatch between phoneme sequences and speech sequences. The input is speech and text pairs passed to a speech encoder and a phoneme encoder. Both representations are connected in joint multimodal space. The space is learned with the frame-level (dis)similarity of speech and text pairs in a batch using contrastive learning. As the extracted intermediate representation contains contextual information, only the current frame corresponds to a positive sample.

$\hat{S}$ denotes the input batch of speech data, $\hat{S} \in \mathbb{R}^{B \times T_s \times D_s}$ where $B$ is the batch size, $T_s$ is the number of time bins, and $D_s$ is the number of spectral components (e.g. Mel bins). $\hat{P}$ denotes the input batch of phoneme data, $\hat{P} \in \mathbb{R}^{B \times T_p \times D_p}$ where $T_p$ is the number of phoneme codes, and $D_p$ is the number of phoneme coding dimensions. During training, the ground-truth duration is used to expand the phoneme sequence's length to $T_s$. During inference, the corresponding predicted duration is used. The phoneme sequence after the length regulator is obtained: $\hat{P} \in \mathbb{R}^{B \times T_s \times D_p}$. ${S} \in \mathbb{R}^{B \times T_s \times d}$ are the speech representations, and ${P} \in \mathbb{R}^{B \times T_s \times d}$ are the phoneme representations. We bring speech and phoneme representations, ${S}$ and ${P}$, into a joint multimodal space of dimension $d$. 

Both $S$ and$P$ are input to the same $Decoder$ with prompt embedding $G$ to predict mel-spectrograms. The predicted mel-spectrograms are compared with the ground-truth mel-spectrograms using mean square error (MSE) loss. The total loss of the model is: $\mathcal{L} = \mathcal{L}_{contrastive} + \mathcal{L}_{mse} $

\subsection{Conditional Diffusion Model}

The acoustic diffusion model calculation is shown in Algorithms \ref{alg:training} and \ref{alg:sampling}. The model uses $q(data), x_0, s, t$, and $p$ to represent data distribution, acoustic coding, semantic coding, diffusion step, and prompt embedding, respectively. One notable feature of the model is that it allows for closed-form sampling of $x_t$ at any timestep $t$ using $\bar{\alpha}_t$ and $\alpha_t$. The non-autoregressive network $\epsilon_{\theta}$ predicts $\epsilon$ from $x_t, t, p$, and $s$. The training objective is to minimize the unweighted variant of the ELBO\cite{ho2020denoising}, as shown in line 7 of Algorithm \ref{alg:training}. The sampling process is shown in Algorithm \ref{alg:sampling}, where $x_T \sim \mathcal{N}(0, I)$ is first sampled, followed by sampling $x_{t-1}\sim p_{\theta}(x_{t-1}|x_t)$ for $t=T, T-1,\cdots,1$. The output $x_0$ is the sampled data. 

The acoustic diffusion model uses a bidirectional dilated convolution architecture with $N$ residual layers grouped into $m$ blocks, each containing $n = \frac{N}{m}$ layers. The dilation is doubled at each layer within each block. Skip connections from all residual layers are summed up, similar to WaveNet\cite{oord2016wavenet}. The model takes in semantic and prompt embeddings as conditional information. The semantic embedding is input to the transformer encoder, upsampled by the length regulator, and added as a bias term for the dilated convolution in each residual layer. The prompt embedding and diffusion step embedding are broadcast over length and added to the input of each residual layer.

The other diffusion-based modules have similar structures but differ in input, diffusion-step, and conditional information. Fig. \ref{fig:proposed_model} illustrates that the duration diffusion model is conditioned on the phoneme sequence. The semantic diffusion model is conditioned on the phoneme sequence upsampled by duration. The wave diffusion model is conditioned on the mel-spectrogram upsampled by frame length.

\algrenewcommand\algorithmicindent{0.5em}%
\begin{figure}[t]
\begin{minipage}[t]{0.5\textwidth}
\begin{algorithm}[H]
  \caption{Training} \label{alg:training}
  \setlength{\belowdisplayskip}{100pt}
  \small
  \begin{algorithmic}[1]
    \Repeat
      \State $x_0, s \sim q(data)$
      \State $t \sim \mathrm{Uniform}(\{1, \dotsc, T\})$
      \State $p = {\hat{\mu}} + {\hat{\sigma}} \odot \phi ; \phi \sim \mathcal{N}(0, I)$
      \State $\varepsilon \sim \mathcal{N}(0, I)$
      \State $\bar{\alpha}_t = \prod_{i=1}^t\alpha_i$
      \State Take gradient descent step on
        \Statex $\nabla _\theta \left\| \varepsilon - \varepsilon_\theta((\sqrt{\bar\alpha_t} x_0 + \sqrt{1-\bar\alpha_t}\varepsilon), t, p, s) \right\|^2$
    \Until{converged}
  \end{algorithmic}
\end{algorithm}
\end{minipage}

\hfill
\begin{minipage}[t]{0.5\textwidth}
\begin{algorithm}[H]
  \caption{Sampling} \label{alg:sampling}
  \setlength{\belowdisplayskip}{100pt}
  \small
  \begin{algorithmic}[1]
    \State $x_T \sim \mathcal{N}(0, I)$
    \For{$t=T, \dotsc, 1$}
      
      \State $\mu_{\theta}(x_t, t, p, s) = \frac{1}{\sqrt{\alpha_t}}\left( x_t - \frac{1-\alpha_t}{\sqrt{1-\bar\alpha_t}} \varepsilon_\theta(x_t, t, p, s) \right)$
      \State $\sigma_{\theta}(x_t, t, p, s) =  \sqrt{\frac{1-\bar{\alpha}_{t-1}}{1-\bar{\alpha}_t}(1-\alpha_t)}$
      
      \State $x_{t-1} = \mu_{\theta} + \sigma_{\theta} \odot \psi; $ $\psi \sim \mathcal{N}(0, I)$ if $t > 1$, else $\psi = 0$
    \EndFor
    \State \textbf{return} $x_0$
  \end{algorithmic}
\end{algorithm}
\end{minipage}
\end{figure}

\section{Experiments}

\subsection{Experimental Step}

 In our experiments, the acoustic diffusion model has 30 residual layers, 64 residual channels, kernel size 3, dilation cycle $[1, 2, \cdots , 512]$, and the linear spaced schedule is $\beta_{t} \in [1\times10^{-4}, 0.05]$ ($T=200$). The other diffusion-based modules have similar structures but differ in diffusion step. The duration diffusion model is $T=5$. The semantic diffusion model is $T=200$. The wave diffusion model is $T=50$. In the CTAP model, the speech encoder consists of 2 convolution layers, a GELU activation function, 6 transformer layers, and a linear layer. The phoneme encoder comprises a convolution layer, a RELU activation function, 4 transformer layers, and a linear layer. The outputs of the speech encoder and phoneme encoder are layer-normalized separately. The hidden dimension of the CTAP is 512. The prompt encoder is a VAE-based model with 6 convolutional layers and a SE-ResNet block. 
 
 The models are trained using 8 NVIDIA TESLA V100 32GB GPUs with a batch size of 12 per GPU. Adam\cite{Kingma2014AdamAM} is used as the optimizer with an initial learning rate of 2e-4.

\subsection{Compared Models and Datasets}

 Due to the inability of standard TTS methods to perform minimally-supervised training, the test sets of {\bf Tacotron-VAE}\cite{qiang2023improving}, {\bf VALL-E}\cite{wang2023neural}, and {\bf NaturalSpeech2}\cite{shen2023naturalspeech} include 3 hours of labeled data from each speaker. In contrast, the minimally-supervised TTS methods {\bf SpearTTS}\cite{zhang2023speak}, {\bf Diff-LM-Speech}\cite{qiang2023minimally}, and {\bf Proposed} use test sets consisting of 15 minutes labeled data and 2.75 hours unlabeled data from each speaker. 
 
 All speech waveforms are sampled at 24kHz and converted to 40-band mel-spectrograms with a frame size of 960 and a hop size of 240. 
 
 To ensure fairness, we modify all methods to utilize the same language model and diffusion model framework. Specifically, the prompt encoder, wave diffusion model, and Grapheme-to-Phoneme (G2P)\cite{qiang2022back} structure are identical across all compared models. Hubert\cite{hsu2021hubert} is used as a control group for semantic coding.

\begin{table}[]
\captionsetup{skip=20pt} 
\vspace{20pt}
 \caption{Prosody Measurement \& WER}
 \label{tab:prosody}
 \centering
 \resizebox{\linewidth}{!}{ 
\begin{tabular}{lccc}
\hline
\multicolumn{1}{c}{Model}                                        & MSEP          & MSED          & WER          \\ \hline
Tacotron-VAE\cite{qiang2023improving}               & 97.4          & \textbf{18.7} & 7.8          \\ \hline
VALL-E\cite{wang2023neural}                     & 98.6          & 19.5          & 6.1          \\ \hline
NaturalSpeech2\cite{shen2023naturalspeech}      & \textbf{95.9} & 25.1          & 4.5          \\ \hline
SpearTTS(Hubert)\cite{zhang2023speak}           & 110.5         & 19.0          & 8.5          \\ \hline
Diff-LM-Speech(Hubert)\cite{qiang2023minimally} & 107.2         & \textbf{18.7} & 7.2          \\ \hline
Proposed(Hubert)                                                 & 103.5         & 20.1          & 4.6          \\ \hline
Proposed(CTAP)                                                   & 98.2          & 20.0          & \textbf{4.4}  \\ \hline
\end{tabular}
}
\end{table}

\begin{table}[]
\captionsetup{skip=20pt} 
\vspace{20pt}
 \caption{Mean Opinion Score (MOS)}
 \label{tab:mos}
 \centering
\resizebox{\linewidth}{!}{ 
\begin{tabular}{lccc}
\hline
\multicolumn{1}{c}{Model} & Prosody Sim           & Speaker Sim           & Speech Quality        \\ \hline
Tacotron-VAE              & 3.82 ± 0.072          & 3.92 ± 0.087          & \textbf{4.01 ± 0.023}  \\ \hline
VALL-E                    & 3.64 ± 0.050          & 3.70 ± 0.052          & 3.61 ± 0.013          \\ \hline
NaturalSpeech2            & 3.73 ± 0.054          & \textbf{4.04 ± 0.086}          & 3.79 ± 0.070          \\ \hline
SpearTTS(Hubert)          & 3.60 ± 0.059          & 3.68 ± 0.030          & 3.50 ± 0.081          \\ \hline
Diff-LM-Speech(H)         & 3.80 ± 0.090          & 3.71 ± 0.015          & 3.88 ± 0.042          \\ \hline
Proposed(Hubert)          & 3.89 ± 0.013          & 3.71 ± 0.077          & 3.83 ± 0.002          \\ \hline
Proposed(CTAP)            & \textbf{3.91 ± 0.010} & 3.94 ± 0.099 & \textbf{4.01 ± 0.020}         \\ \hline
\end{tabular}
}
\end{table}

\subsection{Test Metrics}
We conduct all subjective tests using 11 native judgers, with each metric consisting of 20 sentences per speaker. 
The test metrics used in the evaluation include prosody measurement, which involves mean square error for pitch ({\bf MSEP}) and duration ({\bf MSED}) to assess prosody similarity against ground-truth speech, word error rate ({\bf WER})(200 sentences per speaker), which utilizes an ASR model to transcribe the generated speech, and mean opinion score ({\bf MOS}), which verifies speech quality and similarity in expected speaking prosody and timbre between source speech and synthesized speech.

\subsection{Results}
Table \ref{tab:prosody} shows that one-stage models, such as {\bf Tri-Diff-Speech}, {\bf NaturalSpeech2}, {\bf VALL-E}, and {\bf Tacotron-VAE}, outperform two-stage models like {\bf Spear-TTS}, {\bf Diff-LM-Speech}, and {\bf Proposed(Hubert)} in terms of MSEP. This is because existing semantic representation models like Hubert suffer from information redundancy and dimension explosion, making overall task modeling more challenging than one-stage modeling. On the other hand, our proposed method, {\bf Proposed(CTAP)}, uses an intermediate semantic representation that is independent of paralinguistic information and solves the problem of difficult prediction with a pre-trained frozen phoneme encoder. The phoneme and speech representations extracted by the CTAP model based on contrastive learning are very similar, making the prediction task of the semantic diffusion model simple. In terms of MSED, {\bf Diff-LM-Speech} achieves the best results, while {\bf Proposed(Hubert)} and {\bf Proposed(CTAP)} also perform better than {\bf NaturalSpeech2}'s non-autoregressive structure due to their use of the duration diffusion model. Furthermore, compared to other autoregressive structures ({\bf VALL-E}, {\bf Diff-LM-Speech}, etc.), {\bf Proposed} and {\bf NaturalSpeech2} have significant advantages in synthesizing robust speech due to their non-autoregressive structure, which has been demonstrated in the WER results.

Table \ref{tab:mos} shows that our proposed method outperforms others in terms of prosody similarity MOS. This is because more randomness is introduced into the duration diffusion model, achieving diversified prosodic expression. Among them, {\bf Proposed(CTAP)} has the best results. For speaker similarity MOS, {\bf Proposed(CTAP)} and {\bf NaturalSpeech2} using non-autoregressive structures perform well. Additionally, compared to models that use discrete acoustic coding, all models that use mel-spectrogram as acoustic features achieve better speech quality MOS scores. Specifically, {\bf Proposed(CTAP)} and {\bf Tacotron-VAE} achieve the best results, highlighting the importance of continuous acoustic features for speech quality.

Despite the promising results, our method's biggest problem is that the inference speed is slow due to multiple diffusion models. We plan to address this issue in future work. Initially, we chose a solution that was based entirely on the diffusion module to consider the novelty and consistency of the model. However, as described in the paper, thanks to the powerful modeling ability of the diffusion model, the Duration Diffusion Model, Semantic Diffusion Model, and Acoustic Diffusion Model performed better than other structures. As for the Wave Diffusion Model, its improvement relative to GAN was minimal in our experiments and consumed more computing resources. This is also the focus of our future work.

\section{Conclusions and future work}

We propose a high-fidelity speech synthesis method based on diffusion models with minimal supervision. We introduce the CTAP method as an intermediate semantic representation and use mel-spectrograms as acoustic representations. Our approach addresses issues with existing methods, such as information redundancy, waveform distortion, and instability. Experimental results demonstrate effectiveness. Future work will focus on improving the slow inference speed caused by multiple diffusion models.

\section{Acknowledgments}
This work is supported by the National Natural Sciencel Foundation of China (No. U23B2053, 62101553).

\vfill\pagebreak

\bibliographystyle{IEEEbib}

\footnotesize
\bibliography{strings,refs}

\end{document}